\documentclass[%
 reprint,
 showkeys,
 amsmath,amssymb,
 aps,
floatfix,
]{revtex4-2}

\usepackage{graphicx}
\usepackage{dcolumn}
\usepackage{bm}

\usepackage[hidelinks]{hyperref}

\newcommand{\NN}{\text{N}\ensuremath{_2^+}}
\newcommand{\Ca}{Ca$^+$}
\newcommand{\ket}[1]{\ensuremath{\left| #1 \right\rangle}}

\newcommand{\NNdown}[0]{\ket{\downarrow}\ensuremath{_{\text{N}_2}}}
\newcommand{\NNup}[0]{\{\ket{\uparrow}\ensuremath{_{\text{N}_2}}\}}
\newcommand{\Cadown}[0]{\ket{\downarrow}\ensuremath{_{\text{Ca}}}}
\newcommand{\Caup}[0]{\ket{\uparrow}\ensuremath{_{\text{Ca}}}}

\begin{document}

\preprint{APS/123-QED}

\title{Molecular-ion quantum technologies}

\author{Mudit Sinhal}
\author{Stefan Willitsch}%
\email{stefan.willitsch@unibas.ch}
\affiliation{%
Department of Chemistry, University of Basel, \\ Klingelbergstrasse 80, 4056 Basel, Switzerland.
}%

\date{\today}

\begin{abstract}
Quantum-logic techniques for state preparation, manipulation and non-destructive interrogation are increasingly being adopted for experiments on single molecular ions confined in traps. The ability to control molecular ions on the quantum level via a co-trapped atomic ion offers intriguing possibilities for new experiments in the realms of precision spectroscopy, quantum information processing, cold chemistry and quantum technologies with molecules. The present article gives an overview of the basic experimental methods, recent developments and prospects in this field.
\end{abstract}

\maketitle
\thispagestyle{myheadings}
\tableofcontents

\section{\label{sec:Introduction}Introduction}

Over the past years, the development of experimental methods which enable the control of single isolated quantum systems has made impressive progress.
While sophisticated techniques for the cooling and coherent manipulation of atomic systems are now well established and form the basis for their application in modern quantum science including quantum computation, quantum simulation and precision measurements \cite{ladd10a, georgescu14a, acin18b, bruzewicz19a, ludlow15a}, their adaption to molecular systems has remained a persistent challenge. The complexity of the molecular energy-level structure reflecting electronic, vibrational and rotational motions as well as electronic- and nuclear-spin degrees of freedom poses, in general, severe difficulties for their translational cooling as well as the preparation, manipulation and readout of individual quantum states. While direct laser cooling of diatomic molecules like SrF \cite{shuman10a} and CaF \cite{williams18, anderegg18a} as well as polyatomic molecules like H$_2$CO \cite{prehn16a}, CaOH \cite{baum21a}, CaOCH$_3$ \cite{mitra20a} has been demonstrated, the vast majority of molecules cannot be laser cooled because they lack closed optical cycling transitions. Thus, alternative cooling schemes have to be implemented \cite{krems09a}. By the same token, sensitive optical-cycling methods which are used for state readout in atomic systems \cite{leibfried03a, haeffner08a, myerson08a} cannot readily be applied. 

While the additional degrees of freedom offered by molecules, in particular rotations and vibrations, impose challenges for their quantum control, they at the same time provide a rich playground for new applications. Spectroscopic transitions in molecular systems span a wide range of frequency bands from the kHz to the PHz regime many of which are not easily accessible in atomic systems. Within the high density of interacting molecular levels, transitions with excellent coherence properties, which are scarce in atomic systems, can readily be identified. Such transitions are attractive candidates for qubits and for precision measurements \cite{najafian20a}. 
Moreover, molecules offer prospects as novel platforms for tests of fundamental physical concepts such as the electric dipole moment of electron (eEDM) \cite{ho20a, baron14a} and possible parity violation effects at the molecular level \cite{altuntacs18a, crassous03a, quack08a}, for precise determinations of fundamental constants \cite{shelkovnikov08a, schiller05a, kajita14a, kortunov21b, patra20a}, for precision spectroscopic measurements \cite{semeria20a, melosso21a}, for tests of ab-initio calculations and molecular quantum theory \cite{hoelsch19a, alighanbari20a, patra20a}, and for investigations of state- and energy-controlled atom-molecule \cite{jurgilas21a, doerfler19a} and ultracold molecule-molecule \cite{cheuk20a} collisions. Molecules are also being increasingly considered for new frequency standards and clocks \cite{kajita15a, najafian20a, karr14a, schiller14a}. Last but not least, molecules are the building blocks of chemistry giving access to a vast variety of chemical and biochemical phenomena the study of which immensely benefits from precise experimental methods \cite{willitsch17a, heazlewood15a, heazlewood21a}.

In this chapter, we discuss upcoming quantum technologies for the coherent control and manipulation of molecular ions in radiofrequency (RF) traps and their application in various domains of molecular science. Quantum-logic schemes in which single molecular ions are trapped together with single atomic ions in ion traps have recently emerged as general and flexible methods for non-destructive quantum-state manipulation of single molecules. In such experiments, operations that cannot directly be performed on the molecular ion due to its complex energy-level structure can be implemented via the co-trapped atomic ion. This new methodology constitutes a paradigm change in how molecules are interrogated: single molecules are used instead of large ensembles, investigations can be performed in a completely non-destructive and even quantum-non-demolition manner, and dramatic improvements in measurement sensitivity and precision can therefore be realised. As will be reviewed in this chapter, such protocols have so far successfully been employed for the cooling of molecular ions to the motional ground state of the trap, for the preparation and non-destructive detection of molecular quantum states, for extremely sensitive spectroscopic measurements and for the implementation of coherent operations \cite{sinhal20a, najafian20b, meir19a, wolf16a, chou17a, chou20a, lin20a}, while a variety of further applications has been proposed. 

We begin by discussing the basic concepts and methods used in ion trapping and cooling. To contrast with recent developments, we briefly summarise destructive techniques that have been used in the past to probe trapped molecular ions. We then give an overview of the most recent developments in molecular-ion quantum technologies with a focus on advances in the non-destructive manipulation and coherent control of single molecular ions. Finally, we conclude with an outlook on future developments in the field.

\section{Experimental Techniques}

\subsection{Ion trapping}

The experiments discussed here all rely on the confinement of ions in RF traps, also referred to as Paul traps. 
The theory of RF traps has been extensively treated in text books and review articles, see, e.g., Refs. \cite{major05a, march05a, leibfried03a, drewsen00a, knoop14a, thompson16a, ghosh95a}. Here, we briefly discuss the principles of linear-quadrupole RF traps (LQTs) which are of particular relevance in the present context.  

\begin{figure}[t!] 
    \centering
    \includegraphics[width=0.9\linewidth,trim={0cm 0cm 0cm 0cm},clip]{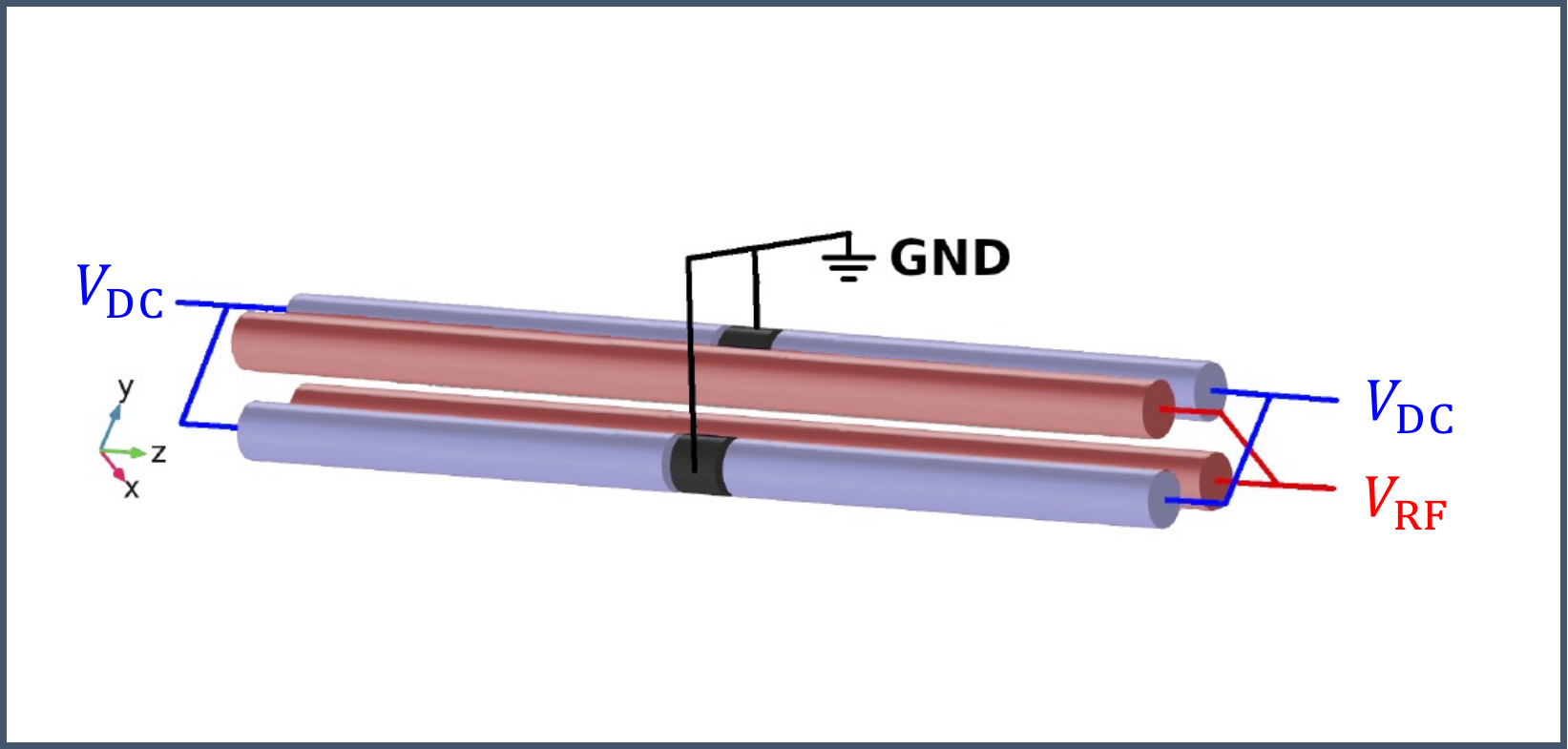}
    \caption[Schematic representation of a linear Paul trap]{Schematic representation of a linear-quadrupole radiofrequency ion trap consisting of four cylindrical electrodes arranged in a quadrupolar configuration. Ions are confined in the centre of the trap by applying radiofrequency voltages $V_{\text{RF}}$ and static voltages $V_{\text{DC}}$ to the electrodes as indicated.}
    \label{fig:IonTrapSchematic}
\end{figure}

Figure \ref{fig:IonTrapSchematic} shows a schematic of a generic linear Paul trap consisting of four cylindrical electrodes arranged in a quadrupolar configuration. Note that LQTs can be configured in a variety of three- and two-dimensional electrode geometries including cylinders, wires, blades, wafers and surface structures \cite{knoop14a, thompson16a}. Because Earnshaw’s theorem forbids the generation of an electrostatic potential minimum, ion traps employ RF fields in order to confine ions dynamically. A dynamic trapping field in the radial ($x$, $y$) directions is generated by applying a time-varying voltage $V_{\text{RF}} = V_0 \cos(\Omega_{\text{RF}}t)$ to a pair of diagonally opposed electrodes (shown in red in Fig. \ref{fig:IonTrapSchematic}). Spatial confinement of the ions along the axial ($z$) direction is provided by applying static potentials, $V_{\text{DC}}$, on the endcaps of the trap (shown in blue in Fig. \ref{fig:IonTrapSchematic}).

An LQT is designed to generate an approximately quadrupolar electric potential close to the trap centre. The equations of motion for a single ion in such a trap take the form of Mathieu equations \cite{march97a, major05a, drewsen00a},

\begin{equation} \label{eqn:mathieu_eqn}
    \frac{d^2 u}{d\tau^2} + \left[ a_u - 2q_u\cos(2\tau) \right] u  = 0,
\end{equation}
where $u \; \in \;  \{x, y, z\}$. The Mathieu parameters, $a_u$ and $q_u$, are given by, 
\begin{subequations} \label{eqn:mathieu_params}
\begin{align}
    a_x &= a_y = -\frac{1}{2}a_z = -\kappa \, \frac{4eV_{\text{DC}}}{m z_0^2\Omega_{\text{RF}}^2},
    \\
    q_x &= -q_y = \frac{2 e V_{\text{RF}}}{m r_0^2\Omega_{\text{RF}}^2}, \; \; q_z = 0, \intertext{and,}
    & \qquad \quad \tau = \frac{\Omega_{\text{RF}}t}{2},
\end{align}
\end{subequations}

where, $m$ and $e$ are the mass and the charge of the trapped ion, respectively, 2$z_0$ is the distance between the endcaps, 2$r_0$ is the distance between two diagonally opposed electrodes and $\kappa$ is a geometry parameter describing the stiffness of the static axial trapping potential which is considered to be harmonic \cite{drewsen00a}. Stable trapping of ions is possible for certain ranges of the $a_u$ and $q_u$ parameters which is usually represented in terms of stability diagrams \cite{drewsen00a}. 

In the limit of $|a|$, $|q|$ $\ll$ 1, the motion of a trapped ion according to Eq. \ref{eqn:mathieu_eqn} can approximately be described in terms of two superimposed components, a slow, `secular’ harmonic motion with frequency $\omega_u$ and a fast `micromotion' at the frequency  $\Omega_{\text{RF}}$ of the RF drive. The trajectory of the ion can thus be expressed as \cite{berkeland98a}, 
\begin{equation} \label{eqn:ion_trajectory}
    u(t) \approx A_u \cos(\omega_u t + \phi) \left[ 1 - \frac{q_u}{2}\cos(\Omega_{\text{RF}}t) \right],
\end{equation}
where $\phi$ is a phase describing the ion's initial position and velocity and $A_u \equiv A_x$, $A_y$, $A_z$ are the amplitudes of the motion. In Eq. \ref{eqn:ion_trajectory}, the first term corresponds to the secular and the second term to the micromotion. The secular motional frequencies are given by,

\begin{equation}
    \omega_u = \frac{\Omega_{RF}}{2} \sqrt{a_u + \frac{q_u^2}{2}}.
\end{equation}

In the quantum limit, the secular motion of the ion can be represented by a quantum harmonic oscillator with discrete motional states \ket{n_u} characterised by motional quantum numbers $n_u = 0$, $1$, $2$, $3$, ... and energies $E_{n_u} = (n_u + 1/2)\hbar\omega_u$. For an ion in the motional ground state \ket{n_u = 0}, the extent of its wavefunction is given by, 
\begin{equation}
    u_{0} = \sqrt{\frac{\hbar}{2m\omega_u}}.
\end{equation}
In the interaction of the ion with light, the quantity $\eta = ku_{0}$ (the ``Lamb-Dicke parameter") represents the ratio between the wavelength of the radiation with wave vector $k = 2\pi/\lambda$ and $u_{0}$ \cite{leibfried03a}. In the Lamb–Dicke regime, defined as $\eta^2(2\,\bar{n}_u+1) << 1$, where $\bar{n}_u$ represents the average motional quantum number of the ion in the trap, the extent of the ion's wavefunction is much smaller than the wavelength of the radiation coupling to the ion. In this regime, first-order Doppler shifts due to the motion of the ion vanish and transitions between different motional states appear as discrete sidebands modulated onto an ionic spectrum \cite{leibfried03a}. 

\begin{figure}[t!] 
    \centering
    \includegraphics[width=1.0\linewidth,trim={0cm 0cm 0cm 0cm},clip]{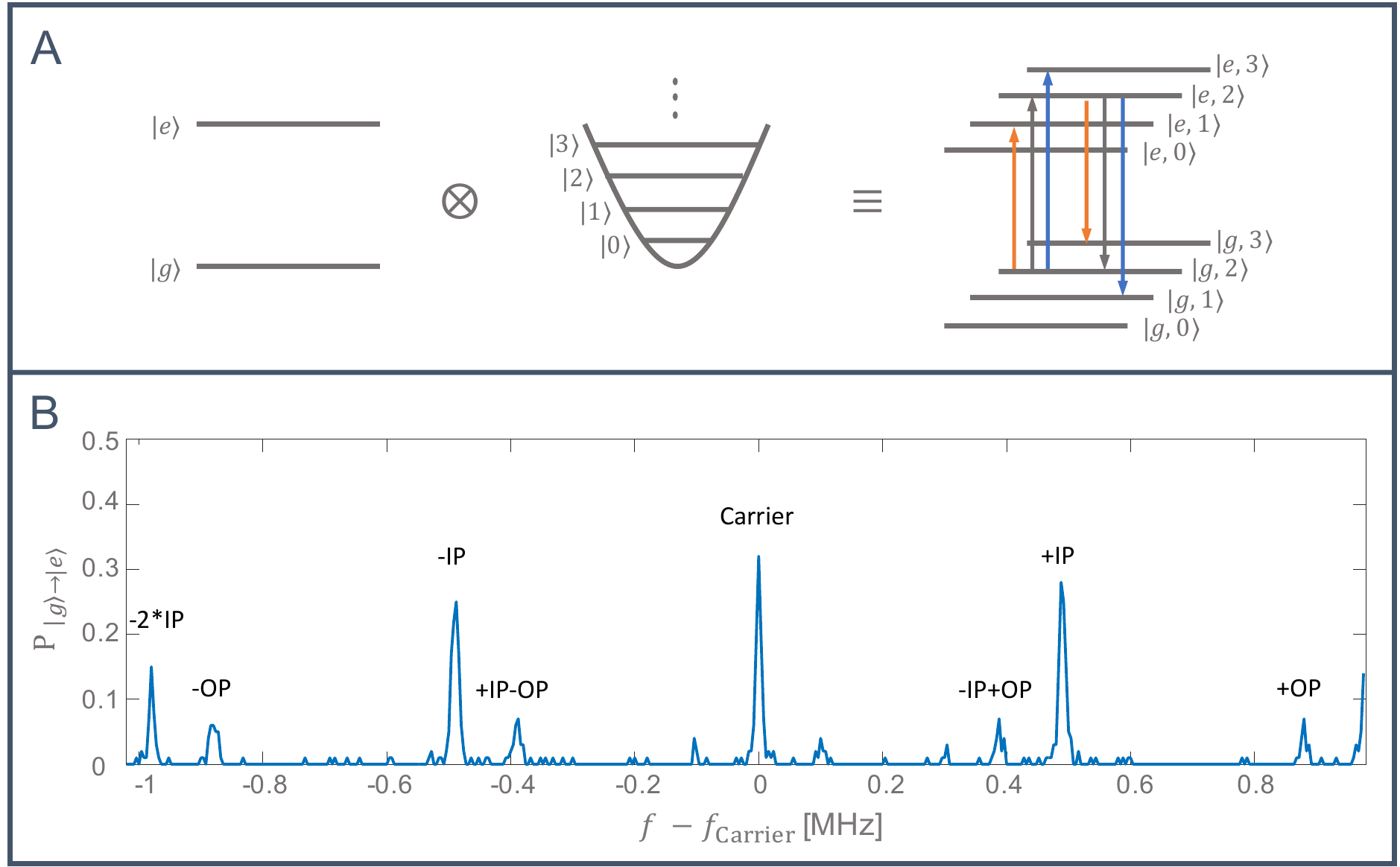}
    \caption[Quantization of energy levels of an ion trapped in a harmonic potential]{{\bf A} Illustration of the combined internal-motional energy levels of a  two-level system trapped in a harmonic potential. In the Lamb-Dicke regime, transitions between the quantised motional states of the particle appear as red and blue sidebands (red and blue arrows respectively) modulated onto carrier transitions (grey arrows). {\bf B} Probability ($\text{P}_{\ket{g} \rightarrow \ket{e}}$) for exciting the $\ket{g} = \text{(4s)}^2\text{S}_{1/2}(m = -1/2) \rightarrow \ket{e} = \text{(3d)}^2\text{D}_{5/2}(m = -5/2)$ transition in \Ca{} (also see Fig. \ref{fig:CaCooling}) for a \Ca{}--\NN{} two-ion Coulomb crystal as a function of the laser frequency, $f$, around the carrier frequency, $f_{\text{Carrier}}$. Red (-) and blue (+) sidebands of the in-phase (IP) and out-of-ohase (OP) axial motional modes of the ion crystal are observed. Reproduced from Ref. \cite{meir19a} with permission from the Royal Society of Chemistry.}
    \label{fig:IonTrap_Basic}
\end{figure}

Figure \ref{fig:IonTrap_Basic}A depicts the combined internal-motional energy levels of a hypothetical two-level ion, with internal ground (\ket{g}) and an excited (\ket{e}) state, trapped in a harmonic potential. As the ion's motional frequencies (of order MHz) are usually much smaller than its internal transition frequencies (hundreds of THz for electronic transition), transitions involving motional energy levels appear as modulations on electronic excitations. Transitions of the form $\ket{g, n_u} \rightarrow \ket{e, n_u}$ or $\ket{e, n_u} \rightarrow \ket{g, n_u}$ in which the ion is excited or de-excited without a change in the motional quantum number are termed as carrier transitions (grey arrows in Fig. \ref{fig:IonTrap_Basic}A). Transitions in which the motional quantum number of the trapped ion changes are known as sideband transitions. When a sideband appears at a higher frequency than the carrier (blue arrows), i.e., $\ket{g, n_u} \rightarrow \ket{e, n_u + 1}$ or $\ket{e, n_u} \rightarrow \ket{g, n_u - 1}$, the transition is termed a blue sideband (BSB). By contrast, a transition of the form $\ket{g, n_u} \rightarrow \ket{e, n_u - 1}$ or $\ket{e, n_u} \rightarrow \ket{g, n_u + 1}$ is termed a red sideband (RSB) featuring a lower frequency than the carrier (red arrows). In Fig. \ref{fig:IonTrap_Basic}A, only first-order sidebands with $\Delta n_u=\pm1$ are indicated, but it should be noted that also higher-order sidebands with $|\Delta|n_u>1$ exist \cite{leibfried03a}.

In the context of experiments with multiple trapped ions, the Coulomb interaction between the charged particles couples the motion of the individual ions. The collective motions of a string of ions is then described in terms of normal modes \cite{james98b, morigi01a}. For quantum-logic experiments with two ions of mass $m_1$ and $m_2$, the axial secular motion of the two-ion string can be described by an in-phase (IP) mode with a frequency $\omega_-$ and an out-of-phase mode (OP) with a frequency $\omega_+$ given by \cite{morigi01a}, 
\begin{equation}
\omega_{\mp}^2 = \frac{\omega_z}{m_1} \left( 1 + \frac{1}{M_{12}} \mp \sqrt{1 + \frac{1}{M_{12}^2} - \frac{1}{M_{12}} }\: \right),    
\end{equation}
where $\omega_z$ is the axial secular frequency of the ion with the higher mass, $m_1$, and $M_{12} = m_1/m_2$ is the mass ratio of the two ions. Figure \ref{fig:IonTrap_Basic}B shows a typical spectrum of a \Ca{}--\,\NN{} two-ion chain in an ion trap depicting the carrier and the red- and blue-sideband transitions (represented  by ``-'' and ``+'' signs, respectively) corresponding to the IP and OP axial modes.

\subsection{Generation of molecular ions and their state initialization}

\begin{figure*}[tp!] 
	\centering
	\includegraphics[width=1.0\linewidth,trim={0cm 0cm 0cm 0cm},clip]{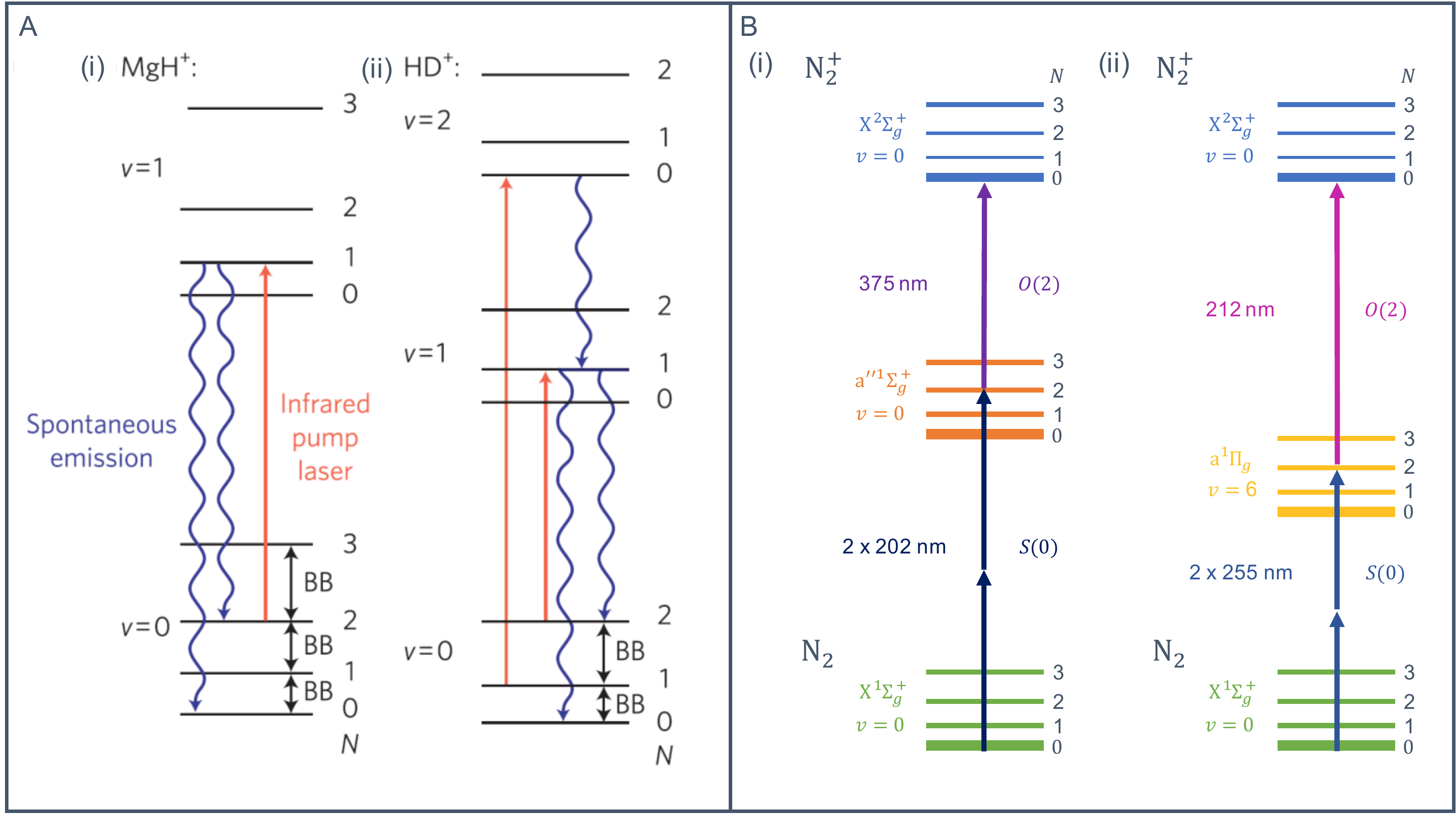}
	\caption[State Preparation]{Rovibrational quantum-state preparation techniques for molecular ions: {\bf A} Trapped (i) MgH$^+$ \cite{staanum10a} and (ii) HD$^+$ \cite{schneider10a} molecular ions are prepared in the rovibrational ground state ($N = 0, v = 0$) by continuously pumping population from selected rotational levels in the ground vibrational state ($v = 0$) to excited levels using infrared lasers (red arrows). The excited molecules can decay to the ground state ($N = 0, v = 0$) by spontaneous emission (blue wavy arrows). Coupling to the ambient blackbody-radiation field (BB; black arrows) leads to a redistribution of the level populations which ultimately accumulate in the rotational ground state. Reproduced from Ref. \cite{willitsch10a}. 
	{\bf B} Two variants of state-selective photoionisation schemes for the preparation of \NN{} molecular ions in the rovibronic ground state \cite{tong10a, gardner19a}. Both schemes start from rotational levels of the vibrationless ($v=0$) $X^1\Sigma^+_g$ ground electronic state, but differ in the choice of intermediate electronic and vibrational states for the resonance-enhanced photoionisation sequence. In (i), the $a''^1\Sigma^+_g, v=0$ state is employed  \cite{tong10a} while in (ii) the $a^1\Pi_g, v=6$ state is chosen  \cite{gardner19a}. In both cases, the wavelength of the ionisation laser is selected so as to only populate the rotational ground state of the cation.}
	\label{fig:StatePrep}
\end{figure*}

For quantum experiments with trapped molecular ions, the clean chemical preparation of the ions in the trap and their initialisation in the required internal state is imperative.
Molecular hydrides like CaH$^+$ and MgH$^+$, which are frequently used species in the present context, are usually produced by chemical \cite{rugango16a, calvin18a, chou17a, chou20a, lin20a, wolf16a} or photo-chemical reactions \cite{molhave00a} of the relevant laser-cooled alkaline-earth ions with hydrogen gas lekaed into the vacuum chamber. Homonuclear ions like O$_2^+$, \NN{}, H$_2^+$ and their isotopomers can be generated by photoionisation \cite{dochain15a, schmidt20a, tong10a, tong11a, gardner19a} or electron-impact ionization \cite{blythe05a} of their neutral precursor molecules. 

The initialisation of the molecular ions in specific internal, i.e., electronic, vibrational, rotational and even hyperfine-Zeeman, states (usually the absolute ground state) can be achieved by a variety of methods. Polar molecular ions whose rotational-vibrational (rovibrational) degrees of freedom couple to dipole radiation can be prepared in specific rovibrational levels by blackbody-radiation-assisted optical pumping techniques \cite{staanum10a, schneider10a, wolf16a, chou17a}, see Fig. \ref{fig:StatePrep}A for examples. Extensions of these schemes have also been used for hyperfine-state preparation \cite{bressel12a}. As a further development, broadband optical-pumping schemes using femtosecond lasers which address multiple rotational transitions simultaneously have been shown to achieve state preparation on considerably reduced timescales \cite{lien14a}. 


By contrast, homonuclear diatomic ions like \NN{} and O$_2^+$ do not feature a permanent electric dipole moment. Consequently, their rovibrational degrees of freedom do not couple to dipole radiation which renders direct optical-pumping schemes inefficient. As an alternative, state-selective photoionisation schemes can be implemented for producing the ions in well-defined rotational states. The technique relies on optical selection and propensity rules which lead to the population of a limited number of rotational states in the ion upon photoionisation of its neutral precursor molecule. By exciting the neutral molecule to a well-defined intermediate electronic state in a first step and subsequently setting the photoionization energy slightly above the lowest ionic state accessible due to selection rules (threshold photoionisation), molecular ions could be generated in single rotational levels, see Fig. \ref{fig:StatePrep}B \cite{tong10a, tong11a}.

Finally, a probabilistic state preparation of the molecular ions can also be achieved by the state-readout schemes discussed in Sec. \ref{sec:qls} below.

\subsection{Cooling of trapped ions}

\begin{figure}[tp!] 
    \centering
    \includegraphics[width=0.7\linewidth,trim={0cm 0cm 0cm 0cm},clip]{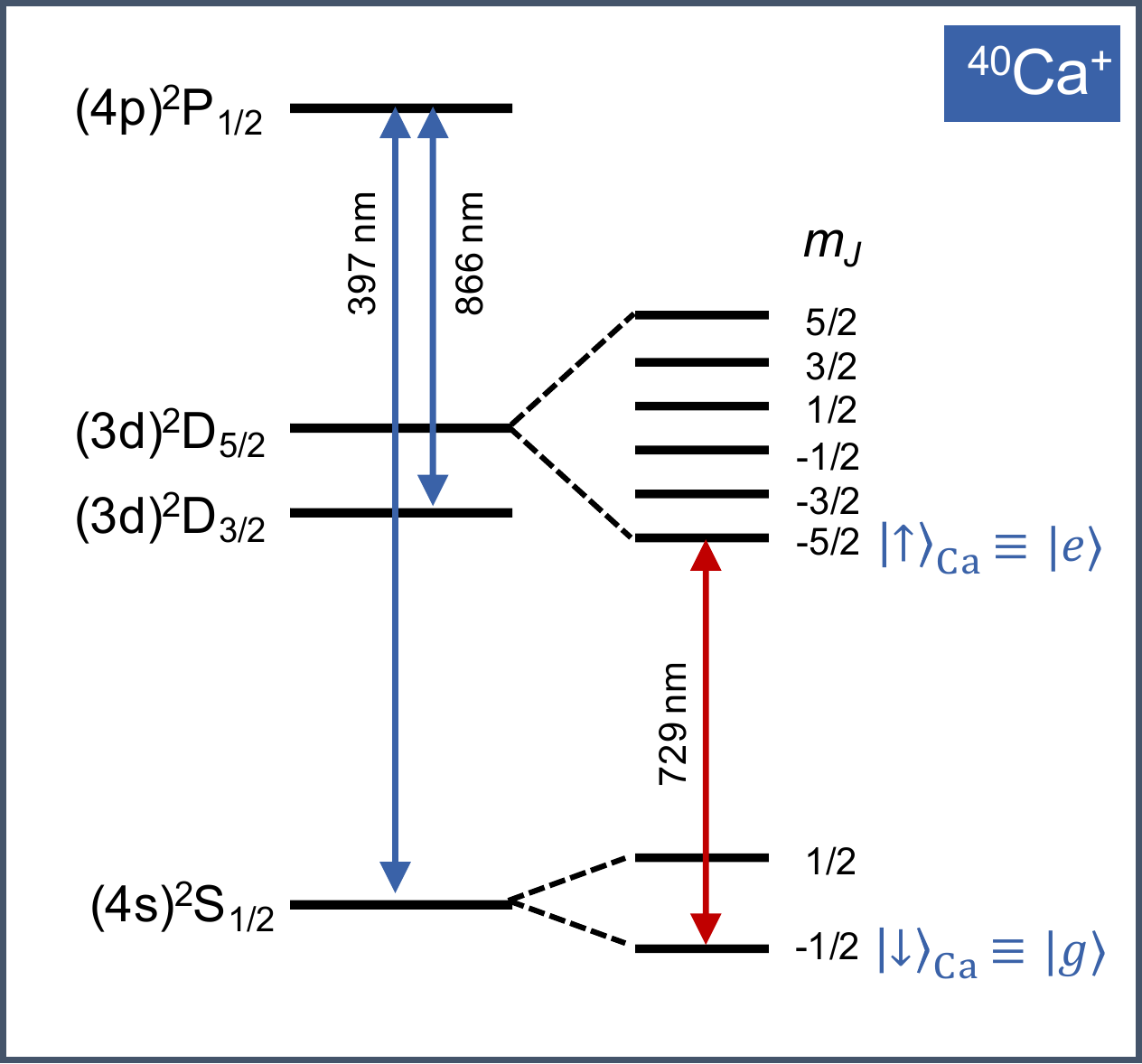}
    \caption[Energy-level structure of the \Ca{} ion]{Energy-level scheme for the laser cooling of $~^{40}$\Ca{} ions. Cooling to temperatures close to the Doppler limit is achieved by the repeated scattering of photons on the $\text{(4s)}^2\text{S}_{1/2} \rightarrow \text{(4p)}^2\text{P}_{1/2}$ transition at 397 nm. A second laser beam at 866 nm is required to repump population from the meta-stable $\text{(3d)}^2\text{D}_{3/2}$ state in order to close the laser cooling cycle. Red sidebands can be addressed repeatedly on, e.g., the narrow $\text{(4s)}^2\text{S}_{1/2}(m = -1/2) \leftrightarrow \text{(3d)}^2\text{D}_{5/2}(m = -5/2)$ electric-quadrupole transition in order to cool the ion to the motional ground state. In this case, an additional re-pumping laser at 854~nm is employed in order to drive the $\text{(3d)}^2\text{D}_{5/2} \rightarrow \text{(4p)}^2\text{P}_{3/2}$ transition (omitted for clarity in the above representation) from which the ion can relax to the ground state by spontaneous emission. Adapted with permission of American Association for the Advancement of Science, from Sinhal et al., \cite{sinhal20a}, 367(6483), 2020; permission conveyed through Copyright Clearance Center, Inc.}
    \label{fig:CaCooling}
\end{figure}

Ions loaded into a trap are typically hot with secular translational temperatures vastly exceeding room temperature. Thus, multi-stage cooling schemes are usually required to prepare the ions in their motional ground state which is the starting point for the advanced experimental protocols discussed in this article.

In a first Doppler-cooling stage, photons are repeatedly scattered from a closed optical transition with a laser frequency slightly detuned to the red from resonance \cite{metcalf99a}. Alkaline-earth ions like Be$^+$, Mg$^+$, \Ca{}, Ba$^+$ and Sr$^+$ exhibit simple energy-level structures which enable the implementation of the required closed optical cycles. As an example, Fig. \ref{fig:CaCooling} depicts the Doppler-cooling scheme employed for \Ca{} ions. As most molecular ions do not possess optical cycling transitions suitable for laser cooling, their motion has to be cooled sympathetically by the interaction with simultaneously trapped, laser-cooled atomic ions \cite{molhave00a}. The limiting temperature achievable by Doppler laser cooling is given by \cite{metcalf99a}, 
\begin{equation}
    T_\text{D} \approx \frac{\hbar \gamma}{2k_\text{B}},
\end{equation}
where $\gamma$ is the natural linewidth of the cooling transition, $\hbar$ is the reduced Planck constant and $k_\text{B}$ is the Boltzmann constant. $T_\text{D}$ is typically on the order of 1 mK, which under typical trapping conditions is usually sufficient for the establishment of the Lamb-Dicke regime. Under these conditions, the ions localise in the trap forming ``Coulomb crystals'' in which individual particles can be observed, addressed and manipulated, see Fig. \ref{fig:DestructiveScheme} \cite{molhave00a, willitsch08a, willitsch12a}.



Following Doppler cooling, a second cooling stage needs to be implemented that repeatedly addresses red-sideband transitions to optically pump the ions into their motional ground state. This can either be direct sideband cooling on a narrow optical transition, e.g., the $\text{(4s)}^2\text{S}_{1/2}(m = -1/2) \leftrightarrow \text{(3d)}^2\text{D}_{5/2}(m = -5/2)$ transition in Ca$^+$ (Fig. \ref{fig:IonTrap_Basic}) \cite{schmidt-kaler03a}, Raman sideband cooling as is frequently used in Be$^+$ \cite{monroe95a}, or electromagnetically-induced-transparency (EIT)-based sideband cooling \cite{morigi00a, schmidt-kaler01a}. In a mixed-species ion chain, sideband cooling in the Lamb-Dicke regime needs to address specific collective modes of the ion motion \cite{home13a}. For many protocols, it is sufficient to cool only one mode of motion to its ground state, e.g., the IP or OP axial mode.



\section{Destructive state-readout techniques}
 

\begin{figure*}[tp!] 
    \centering
    \includegraphics[width=1.0\linewidth,trim={0cm 0cm 0cm 0cm},clip]{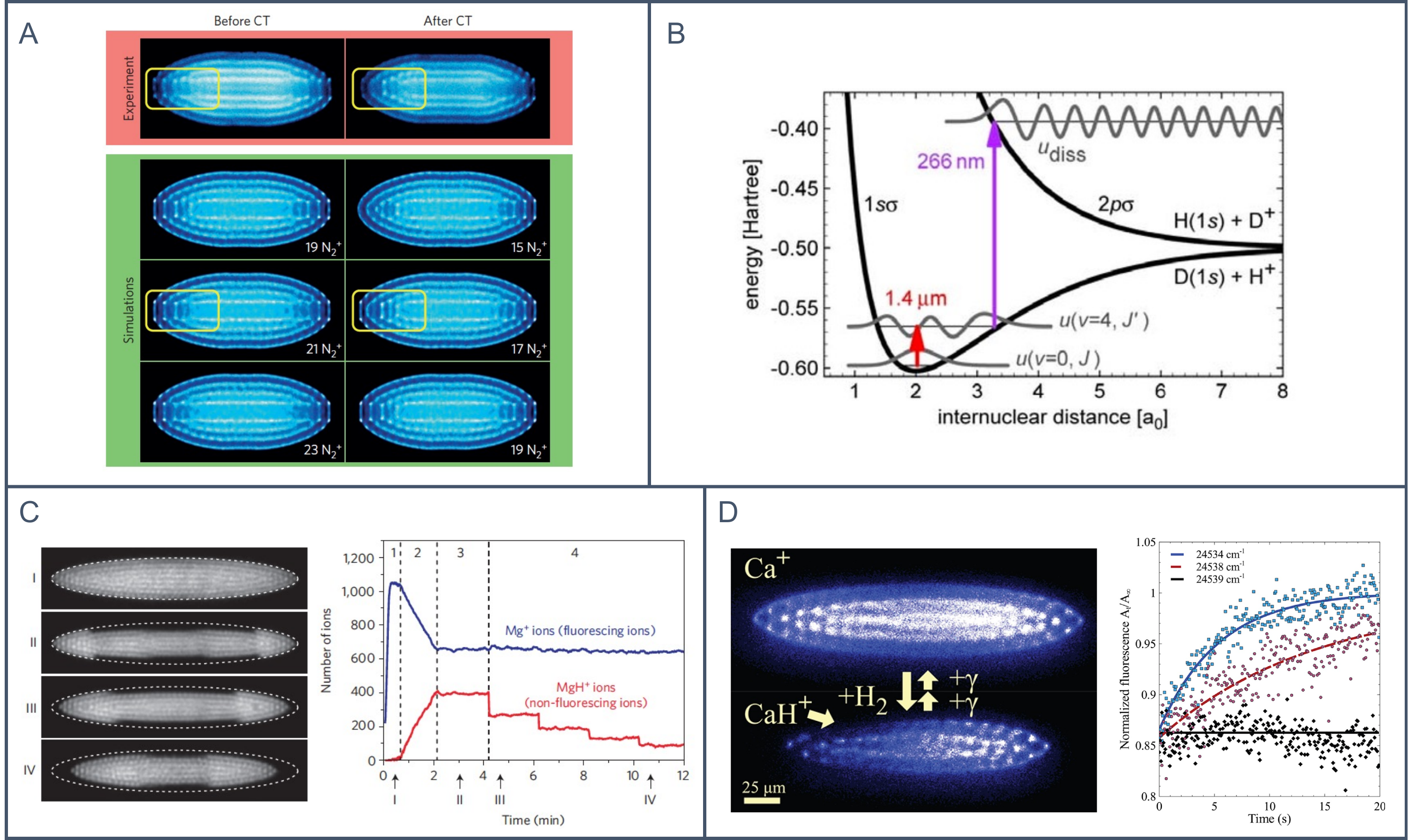}
    \caption[Destructive Schemes]{Destructive state-detection and spectroscopic techniques for trapped molecular ions: {\bf A} Fluorescence images of a Coulomb crystal of \Ca{} and \NN{} ions before and after laser-induced charge transfer (CT) with Ar atoms following their infrared excitation to the first excited vibrational state \cite{germann14a}. The images were obtained by recording the laser-induced fluorescence produced during Doppler cooling of the Ca$^+$ ions. Molecular ions appear as a dark, non-fluorescing region in the centre of the crystal. The number of molecular ions disappearing from the crystal due to CT was found by comparison of the experimental fluorescence images (red panel) with molecular-dynamics simulations (green panel). Reproduced from Ref. \cite{germann14a}.  
    {\bf B} Resonance-enhanced multiphoton dissociation (REMPD) state-detection scheme implemented for rovibrational spectroscopy of HD$^+$ at 1.4~$\mu$m \cite{koelemeij07a}. A single photon at 266 nm was employed to dissociate HD$^+$ ions from the $v = 4$ vibrational state following excitation with the spectroscopy laser. Reprinted figure with permission from \cite{koelemeij07a}. Copyright 2007 by the American Physical Society.
    {\bf C} REMPD of sympathetically cooled MgH$^+$ ions \cite{staanum10a}. The images on the left-hand side show (I) the initial pure Mg$^+$ Coulomb crystal, (II) the crystal after loading MgH$^+$ ions (visible as the non-fluorescing region at the extremities of the crystal), and the crystal after (III) one and (IV) four REMPD cycles. The figure on the right-hand side shows the number of ions deduced from the images as a function of the experiment time. Reprinted by permission from the Springer Nature Customer Service GmbH: Springer Nature for Staanum et al., \cite{staanum10a}, Copyright 2010. 
    {\bf D} REMPD of Coulomb-crystallised CaH$^+$ ions produced by the chemical reaction of laser-cooled Ca$^+$ ions with H$_2$ gas \cite{calvin18a}. The CaH$^+$ ions are  photodissociated thus regenerating the original Ca$^+$ crystal. The figure on the right-hand side shows the total fluorescence yield of the Ca$^+$ ions as a function of the exposure time to the dissociation laser for various laser wavenumbers. Reprinted with permission from \cite{calvin18a}. Copyright 2018 American Chemical Society.}
    \label{fig:DestructiveScheme}
\end{figure*}

Prior to the advent of quantum-logic-spectroscopic readout schemes, various action-spectroscopy techniques have been used in order to probe Coulomb-crystallized molecular ions. As chemical changes of the trapped molecular ions can, in principle, be detected with unit probability using mass spectrometry \cite{drewsen04a}, these methods are highly sensitive and have been widely used in various spectroscopic experiments and for the measurement of state populations.

As an example, laser-induced charge transfer (LICT) was implemented for performing electronic and vibrational spectroscopy of \NN{} molecular ions co-trapped with atomic \Ca{} ions (Fig. \ref{fig:DestructiveScheme}A) \cite{tong10a, tong11a, germann14a}. These experiments capitalised on a state-dependent charge-transfer reaction of \NN{} with Ar atoms according to $\NN{}(v\geq 1) + \text{Ar} \rightarrow \text{N}_2 + \text{Ar}^+$ which only occurs in excited vibrational states of \NN{} with vibrational quantum numbers $v\geq1$ on energetic grounds \cite{schlemmer99a}. Starting from the vibrational ground state, vibrationally excited states were populated either in direct infrared excitation \cite{germann14a} or indirectly via excitation to the first excited electronic state followed by spontaneous emission \cite{tong10a, tong11a}. The latter scheme was used to map out rotational state populations following the generation of the ions by state-selective threshold photoionisation. 

A complementary technique is resonance-enhanced multiphoton dissociation (REMPD) which relies on the dissociation of the molecular ions following their excitation to a suitable intermediate molecular state. 
Such a scheme was employed, for instance, in the high-resolution infrared spectroscopy of trapped HD$^+$ molecular ions sympathetically cooled by laser-cooled Be$^+$ ions \cite{koelemeij07a, biesheuvel16a, patra20a, kortunov21b}, as shown schematically in Fig. \ref{fig:DestructiveScheme}B, and in the rotational spectroscopy of HD$^+$ \cite{alighanbari18a}. The technique was also employed for the read-out of rotational state populations following rotational laser cooling and buffer-gas cooling of sympathetically cooled MgH$^+$ ions \cite{staanum10a, hansen14a} (Fig. \ref{fig:DestructiveScheme}C). REMPD techniques were also used for the vibronic \cite{rugango16a} and rovibronic \cite{calvin18a} spectroscopy of sympathetically cooled CaH$^+$ ions (Fig. \ref{fig:DestructiveScheme}D). 

A major drawback of action-spectroscopic methods is their inherently destructive nature. They rely on a chemical change of the molecular ion following photoexcitation and thus necessitate a complete re-initialisation of the experiment every experimental cycle. The duty cycle of such experiments is therefore low which impairs measurement statistics and, therefore, sensitivity and precision. As a consequence, these experiments typically employ large Coulomb-crystallised ensembles of ions to improve statistics \cite{hansen14a,germann14a,patra20a,kortunov21b}. This invariably entails higher translational temperatures and thus Doppler and other line broadening effects due to ensemble averaging (except in special cases in which spectroscopic experiments can be performed in the Lamb-Dicke regime under these conditions, see, e.g., Ref. \cite{alighanbari18a}). However, in the context of advanced applications such as coherent and ultrahigh-resolution experiments on molecular ions, it is advantageous, and in many cases necessary, to work with single particles. It thus became expedient to develop novel techniques capable of interrogating single molecular ions which do not destroy the molecule and, ideally, also preserve its quantum state.

\section{Quantum-logic experiments on single trapped molecular ions}
\label{sec:qls}

\begin{figure*}[tp!] 
    \centering
    \includegraphics[width=1.0\linewidth,trim={0cm 0cm 0cm 0cm},clip]{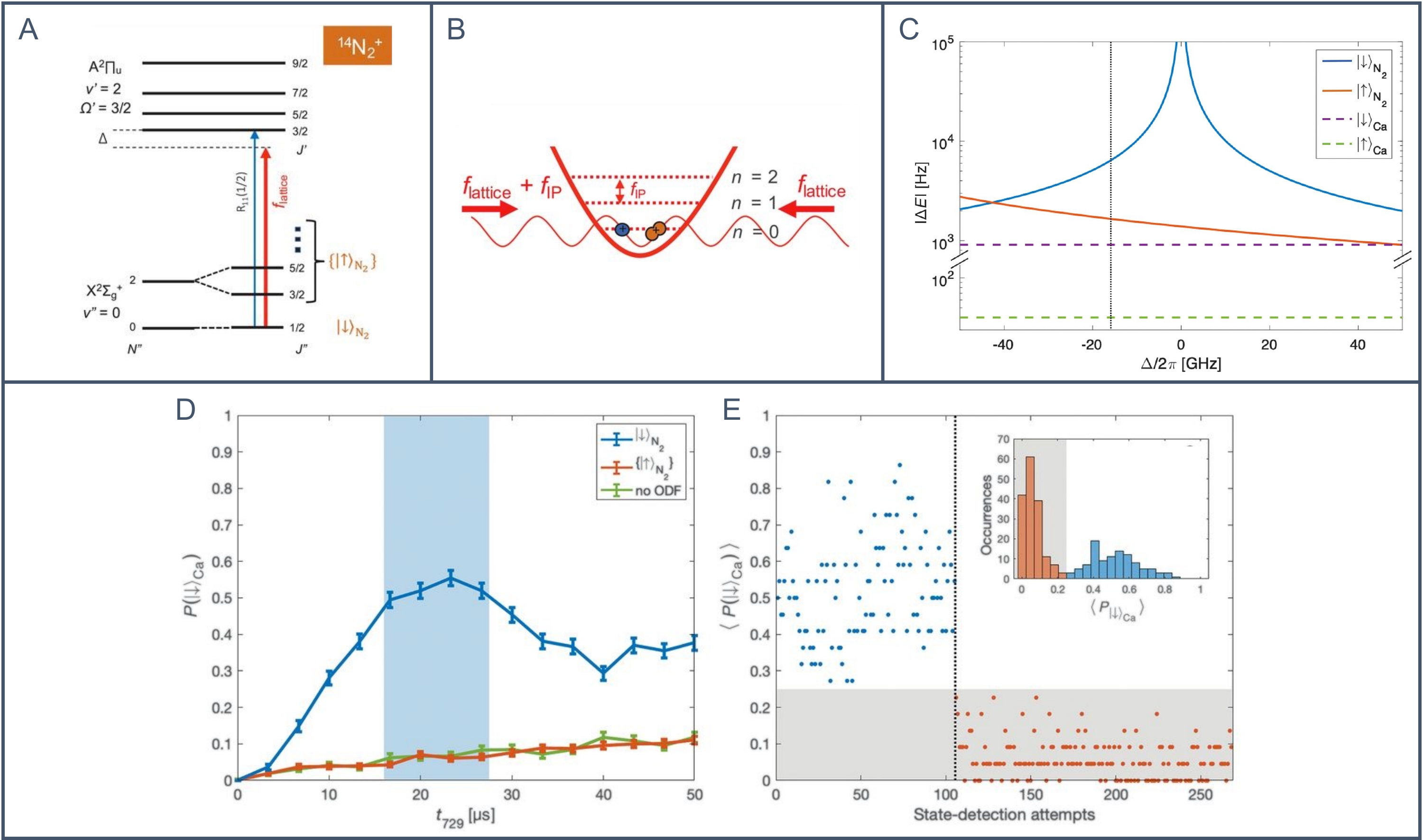}
    \caption[\NN{} Spectroscopy]{Quantum-nondemolition state detection of molecular ions: {\bf A} Simplified energy-level scheme of the \NN{} ion relevant for the quantum-logic-based state determination technique implemented in Refs. \cite{sinhal20a,najafian20b}. {\bf B} Schematic of the protocol implemented for the state readout of \NN{}. Two counter-propagating laser beams form a one-dimensional optical lattice interacting with a \Ca{}-\NN{} two-ion string cooled to the motional ground state of the in-phase (IP) mode in a harmonic trap. One of the lattice beams is detuned by the frequency of the IP mode, $f_{\text{IP}}$, to resonantly excite its motion by an optical dipole force (ODF) depending on the internal state of the \NN{} ion. {\bf C} Calculated magnitude of the ac-Stark shift, $|\Delta E|$, which gives rise to the ODF, experienced by the \NN{} ion as a function of the laser-frequency detuning, $\Delta$, from a strong spin-rovibronic transition in the molecule. At the lattice-laser wavelength indicated by the vertical black dotted line, the \NN{} ion experienced a larger ac-Stark shift in its ground rovibronic state (\NNdown{}) compared to the all other states (\NNup{}) leading to the state selectivity of the detection scheme. In the experiments, the \Ca{} ion was shelved in an excited state \Caup{} in which it experienced a negligible ac-Stark shift (green dashed line) in comparison to the \NN{} ion. Conversely, a \Ca{} ion in its ground state would experience a considerably increased ac-Stark shift (purple dashed line) leading to a spurious background signal in the experiment. {\bf D} Rabi spectroscopy on the \Ca{} \Caup{} $\rightarrow$ \Cadown{} blue-sideband (BSB) transition for \NN{} ion in the \NNdown{} state (blue) and in one of the \NNup{} states (red) as a function of the BSB pulse length $t_{729}$, following coherent motional excitation by the state-dependent ODF. $P(\ket{\downarrow}_{\text{Ca}})$ denotes the population in the $\ket{\downarrow}_{\text{Ca}}$ state. The green trace represents a background measurement of the Rabi oscillation signal without ODF. {\bf E} Time trace of \NN{} state-detection attempts (blue and orange dots) for measuring the molecular state. A threshold of 0.25 (grey shaded area) was used to distinguish between \NN{} in the \NNdown{} or \NNup{} states. A histogram of the state detection attempts is shown in the inset. The figure has been republished with permission of American Association for the Advancement of Science, from Sinhal et al., \cite{sinhal20a}, 367(6483), 2020; permission conveyed through Copyright Clearance Center, Inc.}
    \label{fig:N2_Spec1}
\end{figure*}

Over the past few years, a range of coherent protocols have been developed for manipulating and probing trapped molecular ions. These schemes can be regarded as variants and extensions of the ``quantum-logic spectroscopy'' which was originally conceived in the framework of the Al$^+$ ion optical clock \cite{schmidt05a} and adapted to molecular species. These schemes are inherently non-destructive, thus mitigating the problems associated with the destructive techniques discussed in the previous section, and pave the way for a range of new applications of trapped molecular ions within the realms of quantum science and precision measurements.

In the context of experiments with single trapped molecular ions, we discuss the following three different implementations of quantum-logic protocols. As a first example, we consider the quantum-nondemolition state detection of a single \NN{} ion performed via a co-trapped \Ca{} ion \cite{meir19a, sinhal20a}. The scheme relies on the coherent motional excitation of the ions using molecular-state-dependent optical-dipole forces (ODFs). This approach enabled the tracking of molecular quantum state with a high time resolution as well as rovibronic molecular spectroscopy. In an extension of this scheme \cite{najafian20b}, specific molecular states embedded within a dense energy-level structure could be identified. As a second example, we discuss the non-destructive state detection of MgH$^+$ ions \cite{wolf16a}. In these experiments, an ODF was implemented in order to map the molecular state onto a motional qubit which was then read out on a co-trapped Mg$^+$ ion. As a final example, we present molecular state manipulation schemes implemented for CaH$^+$ ions via co-trapped \Ca{} ions \cite{chou17a, chou20a}. These experiments relied on exciting motional sidebands with Raman transitions on the molecule and allowed for coherent spectroscopy of selected molecular rotational states. Moreover, entanglement between the molecule and the atom could be demonstrated \cite{lin20a}. 


\subsection{Quantum-nondemolition molecular state readout by state-dependent coherent motional excitation}

Fig. \ref{fig:N2_Spec1} illustrates a quantum-non-demolition experiment for the measurement of the spin-rovibronic state of a single \NN{} via a co-trapped  \Ca{} ion \cite{meir19a, sinhal20a}. In the first step of the experimental protocol, the IP mode of the \Ca{}--\NN{} two-ion string was cooled to the ground state of the trap. An optical-dipole force (ODF) the strength of which depended on the molecular state was then used to map the state information onto the motion of the two-ion string. In case the \NN{} was in the state targeted by the ODF (\NNdown{} in Fig. \ref{fig:N2_Spec1} A), the \Ca{}--\NN{} string experienced a large ODF which was used to coherently excite the motion of the previously cooled IP mode. By contrast, for all other molecular states (summarily labeled \NNup{} in Fig. \ref{fig:N2_Spec1}), the magnitude of the ODF was significantly reduced so that only negligible motional excitation occurred. Subsequently, the motional excitation of the two-ion string was detected by Rabi sideband thermometry on the \Ca{} ion thus revealing the information about the internal state of the \NN{} ion. 

The ODF for the state detection was implemented by two counter-propa\-gating laser beams which formed a one-dimensional optical lattice. In order to coherently excite the IP mode of motion, the amplitude of the ODF was modulated by detuning one of the lattice beams by the mode frequency, $f_{\text{IP}}$ (Fig. \ref{fig:N2_Spec1}B). The magnitude of the ODF experienced by the \Ca{}--\NN{} string was dependent on the ac-Stark shift experienced by the \NN{} ion due to the lattice laser. Fig. \ref{fig:N2_Spec1}C depicts the calculated ac-Stark shift, $\Delta E$, for \NNdown{} (here designated as the rovibronic ground state of the molecule) and the \NNup{} states as a function of the lattice-laser detuning from a rovibronic transition (here the $\text{X}^2\Sigma_g^+(v'' = 0) - \text{A}^2\Pi_u(v' = 2)~R_{11}(1/2)$ line \cite{wu07a}, where $v'(v'')$ represent the vibrational  quantum number of the upper(lower) electronic state originating from the \NNdown{} state of \NN{}. The frequency of the lattice laser, $f_{\text{lattice}}$ (black-dotted line), was set such that the a molecule in the \NNdown{} state experienced a large ac-Stark shift and hence a large ODF. Conversely, molecules in excited rotational or vibrational states experienced a much weaker ODF due to the lattice laser being farther detuned from any possible spectroscopic transition originating from these states thus resulting in a smaller motional excitation. For the lattice intensity and polarization chosen in these experiments, a \Ca{} ion in the ground \Cadown{}$=(4\text{s})^2\text{S}_{1/2}(m = -1/2)$ state (Fig. \ref{fig:CaCooling}B) experienced a large, nearly detuning-independent ac-Stark shift (purple-dashed line in Fig. \ref{fig:N2_Spec1}C) leading to a background ODF originating from \Ca{} which impaired the signal-to-noise (SNR) ratio for the detection of the state of the \NN{} ion. Thus, in order to minimize the coupling of the atomic ion to the lattice laser, the \Ca{} ion was shelved in the \Caup{}$=(3\text{d})^2\text{D}_{5/2}(m = -5/2)$ state which experienced a negligible ac-Stark shift (green dashed line) resulting in no appreciable motional excitation due to the \Ca{} ion. 

Fig. \ref{fig:N2_Spec1}D shows the results of Rabi spectroscopy experiments performed on \Ca{} ion after the implementation of the ODF. For \NN{} in the \NNdown{} state, significant motion was excited resulting in Rabi oscillations (blue trace) on the blue sideband of the IP mode of the \Ca{} $\Caup{}\rightarrow \Cadown{}$ transition. By contrast, for molecules in the \NNup{} states, no Rabi oscillations were observed (orange trace) as blue sideband transitions could not be excited when the ion string remained cooled to the ground motional state of the IP mode. Indeed, the signal obtained was comparable to a background measurement of the blue-sideband oscillations on a ground-state cooled \Ca{}--\,\NN{} chain without the application of any ODF (green trace). Each of the traces in Fig. \ref{fig:N2_Spec1}D resulted from thousands of measurements performed on the same molecular ion which was possible due to the non-destructive nature of the state-detection scheme. 

Fig. \ref{fig:N2_Spec1}E depicts the results of the experiments as a function of molecular-state detection attempts. Each data point represents the average of 22 Rabi experiments for which the pulse time was set within the region of maximum contrast indicated by the light-blue area in Fig. \ref{fig:N2_Spec1}D. In this case, a single state-detection attempt took only few 100 ms which allowed for tracking of the molecular state with a high time resolution. For the measurements recorded at the beginning of the experiment (blue dots), the molecule was determined to be in the rovibronic ground state 105 times with zero false events when setting the detection threshold above the grey-shaded area in Fig. \ref{fig:N2_Spec1}E. Subsequently, the molecule changed its state and was detected not to be in the rovibronic ground state 163 times with zero false detection attempts (orange dots). In this experiment, the fidelity for the faithful detection of the rovibronic ground state amounted to $>$99\%.

\begin{figure}[t!] 
    \centering
    \includegraphics[width=0.75\linewidth,trim={0cm 0cm 0cm 0cm},clip]{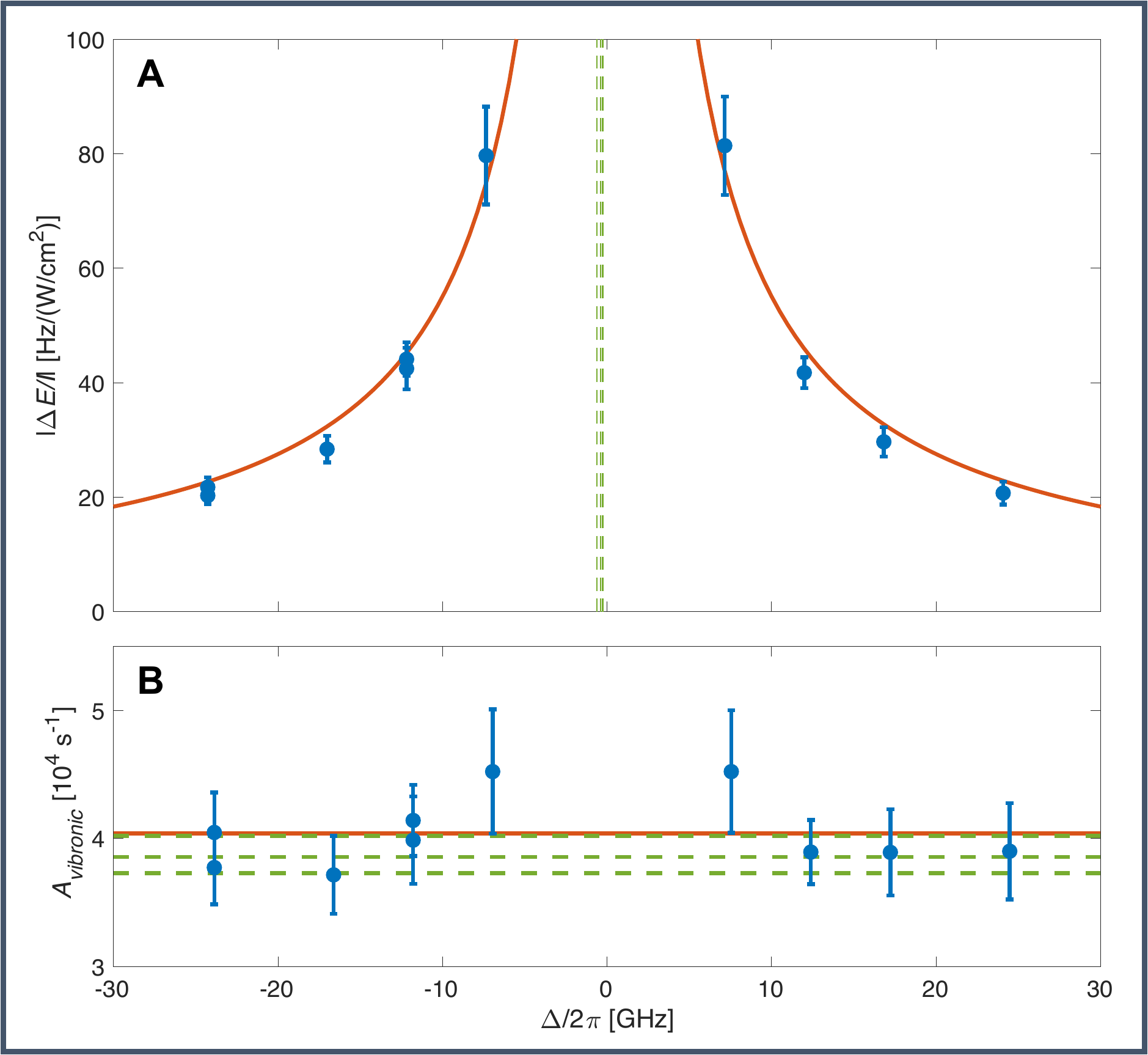}
    \caption[\NN{} Spectroscopy]{Non-invasive molecular spectroscopy of single molecular ions: {\bf A} Line shape of the $\text{X}^2\Sigma_g^+(v'' = 0) - \text{A}^2\Pi_u(v' = 2)~R_{11}(1/2)$ rovibronic transition in \NN{} measured by determining the ac-Stark shift experienced by the ion at various lattice-laser detunings (blue data points) \cite{sinhal20a}. The line center was found by a fit (orange trace) to the measurements in very good agreement with the literature values (green dashed lines). {\bf B} Vibronic Einstein $A$ coefficients, $A_{\text{vibronic}}$, extracted from the measured ac-Stark shifts in panel A at various detunings (blue data points). The mean of the measurements (orange line) was found to be in very good agreement with previous literature values (green dashed lines). Republished with permission of American Association for the Advancement of Science, from Sinhal et al., \cite{sinhal20a}, 367(6483), 2020; permission conveyed through Copyright Clearance Center, Inc. }
    \label{fig:N2_Spec2}
\end{figure}

An application of this state-detection protocol is shown in Fig. \ref{fig:N2_Spec2}. Since the magnitude of the excited motion is dependent on the ac-Stark shift giving rise to the ODF which in turn depends on the detuning of the lattice laser from a spectroscopic transition in the molecule, the scheme can be employed for non-invasive spectroscopy on the molecular ion. By extracting the magnitude of the ac-Stark shift from the observed Rabi oscillations as a function of the lattice-laser detuning, the spectral lineshape of the resonance could be mapped out (Fig. \ref{fig:N2_Spec2}A). From the determined ac-Stark shifts, quantitative spectroscopic parameters like vibronic Einstein-A coefficients, $A_{\text{vibronic}}$, were obtained  (Fig. \ref{fig:N2_Spec2}B). Both line positions and Einstein coefficients were in very good agreement with literature values (green dashed lines in Fig. \ref{fig:N2_Spec2}) thus validating the accuracy of the scheme.

\begin{figure}[tp!] 
    \centering
    \includegraphics[width=0.9\linewidth,trim={0cm 0cm 0cm 0cm},clip]{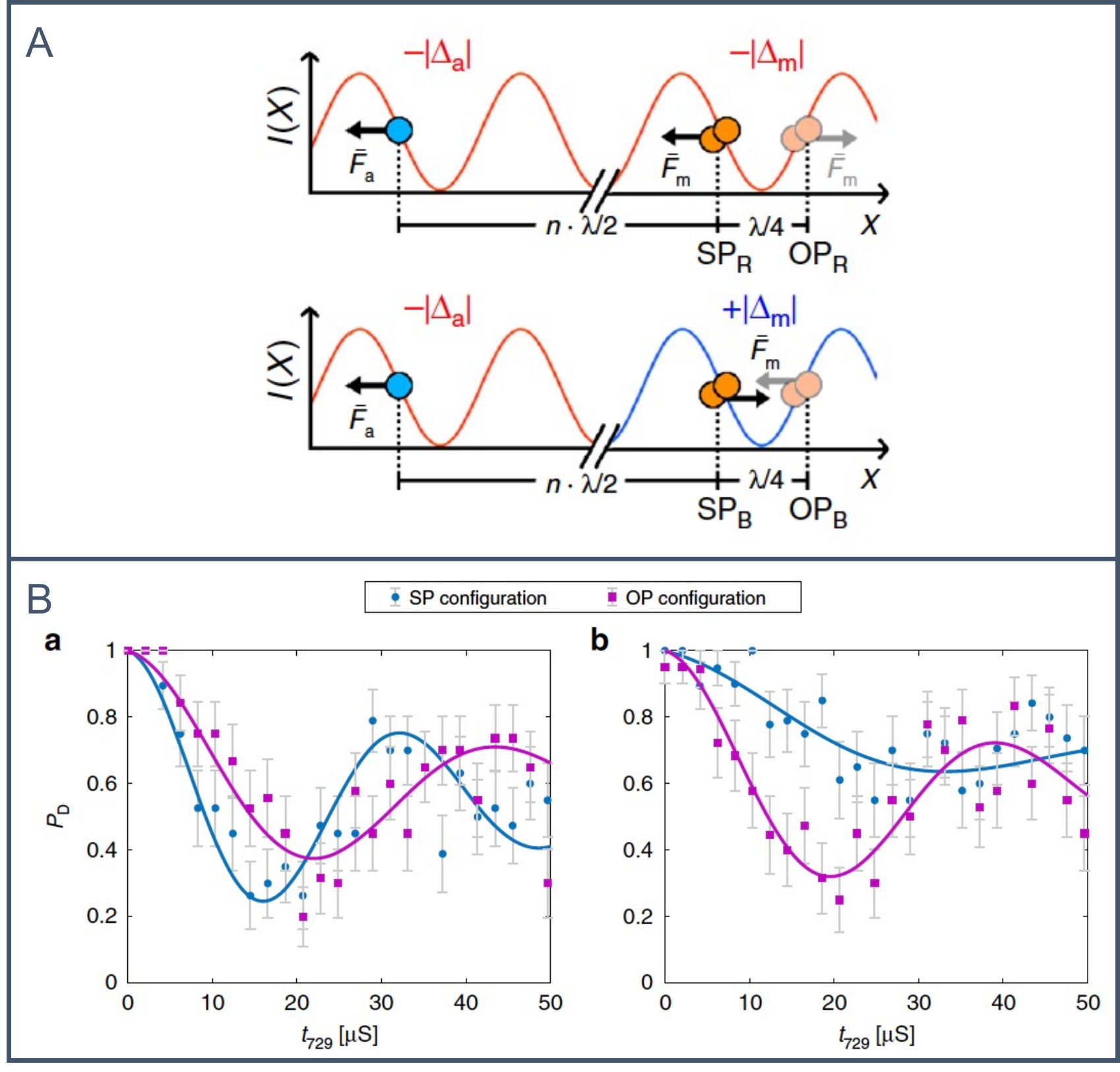}
    \caption[\NN{} Spectroscopy]{ Phase-sensitive forces for molecular state identification. {\bf A} Same-phase (SP) and opposite-phase (OP) configurations of the positions of a \NN{} (orange dots) and \Ca{} (blue dot) ion in an optical lattice in order to measure the amplitude and phase of the ODF experienced by the \NN{} ion. The \Ca{} ion was always red-detuned from its closest resonance and thus always experienced an approximately constant negative ac-Stark shift, $-|\Delta_{\text{a}}|$ and therefore a positive ODF. Depending on the molecular quantum state, the \NN{} ion was either red ($-|\Delta_{\text{m}}|$) or blue ($+|\Delta_{\text{m}}|$) detuned from its nearest resonance resulting in attractive and repulsive optical forces, respectively. {\bf B} If the ODF on \NN{} and \Ca{} have the same sign (sub-panel a), a stronger combined force and therefore stronger Rabi signal is measured on the \Ca{} ion after motional excitation in the SP configuration (blue trace) compared to the OP configuration (magenta trace). Conversely, if the ODF on \NN{} and \Ca{} have opposite signs, the OP configuration gives a stronger Rabi signal (sub-panel b). The relative sign of the ODF on both ions thus yields additional information on the molecular state which is particularly valuable for state-identification within dense energy-levels structures. Reproduced from Ref. \cite{najafian20b}.}
    \label{fig:N2_Spec3}
\end{figure}

In a variation of this scheme, not only the magnitude, but also the sign of the ac-Stark shift were retrieved \cite{najafian20b}. As the ac-Stark shift changes sign across a resonance, its sign encodes the direction of the detuning from resonance (red or blue) and thus valuable additional information for molecular-state identification which is particularly useful in molecules with a dense energy-level structure. Fig. \ref{fig:N2_Spec3} depicts an experiment on a \Ca{}--\,\NN{} chain in which in addition to the \NN{} ion also the \Ca{} ion experienced a finite and constant detuning-independent ODF. An interferometric measurement as shown in Fig. \ref{fig:N2_Spec3}A was then performed in order to detect the molecular state. As before, the measurement started with preparing the \Ca{}--\,\NN{} string in the ground state of the IP mode of motion. The \Ca{} ion was then prepared in the \Caup{} state such that it was always red-detuned from its closest resonance and experienced an ODF in the direction of higher lattice intensity. The \NN{} ion was generated in an arbitrary excited rotational state with the aim to identify this state within the dense Zeeman-spin-rotational energy-level structure in this region. Thus, for \NN{} the amplitude and sign of the ac-Stark shift (and hence the ODF) depended on its specific rovibronic, hyperfine and Zeeman state, while it remained constant and negative for \Ca{}. In order to identify the molecular state, two combined measurements were then performed. In the first experiment, the two ions were placed in the trap such that their separation corresponded to an integer multiple of the lattice nodes. This lattice configuration was denoted as ``same phase'' (SP). In the SP configuration, the ODFs on the molecule and the atom acted in the same (opposite) direction when the molecular ion was red (blue) detuned from its closest resonance (SP$_{\text{R}}$ and SP$_{\text{B}}$ in Fig. \ref{fig:N2_Spec3}A). In the second experiment, the distance between the ions was changed by half a lattice spacing, such that the two ions always experienced an opposite gradient of the lattice field, i.e., the ``opposite phase'' (OP) of the lattice. In the OP configuration, the forces acted in the same (opposite) direction when the molecular ion was blue (red) detuned (OP$_{\text{R}}$ and OP$_{\text{B}}$ in Fig. \ref{fig:N2_Spec3}A). Fig. \ref{fig:N2_Spec3}B depicts Rabi oscillation signals on the \Caup{}$\rightarrow$\Cadown{} BSB transition for the SP (blue) and OP (purple) lattice configurations after an ODF pulse for two different molecular states (panels a and b). In panel a, the stronger SP signal (larger frequency and amplitude of the Rabi oscillation) suggested that the lattice laser was red detuned from the closest molecular transition. By comparing the measured amplitude and sign of the ac-Stark shift with theory, the spin-rotational state of the molecule could be identified as $N=6$, $J=11/2$, where in this case $N$ and $J$ are the rotational and spin-rotational quantum numbers, respectively. In panel b, the stronger OP signal suggested that the lattice laser was blue detuned from the closest molecular transition. In this case, all rotational levels $N\leq4$ could be excluded.


\subsection{Molecular state detection and spectroscopy via a motional qubit}

\begin{figure*}[tp!] 
    \centering
    \includegraphics[width=1.0\linewidth,trim={0cm 0cm 0cm 0cm},clip]{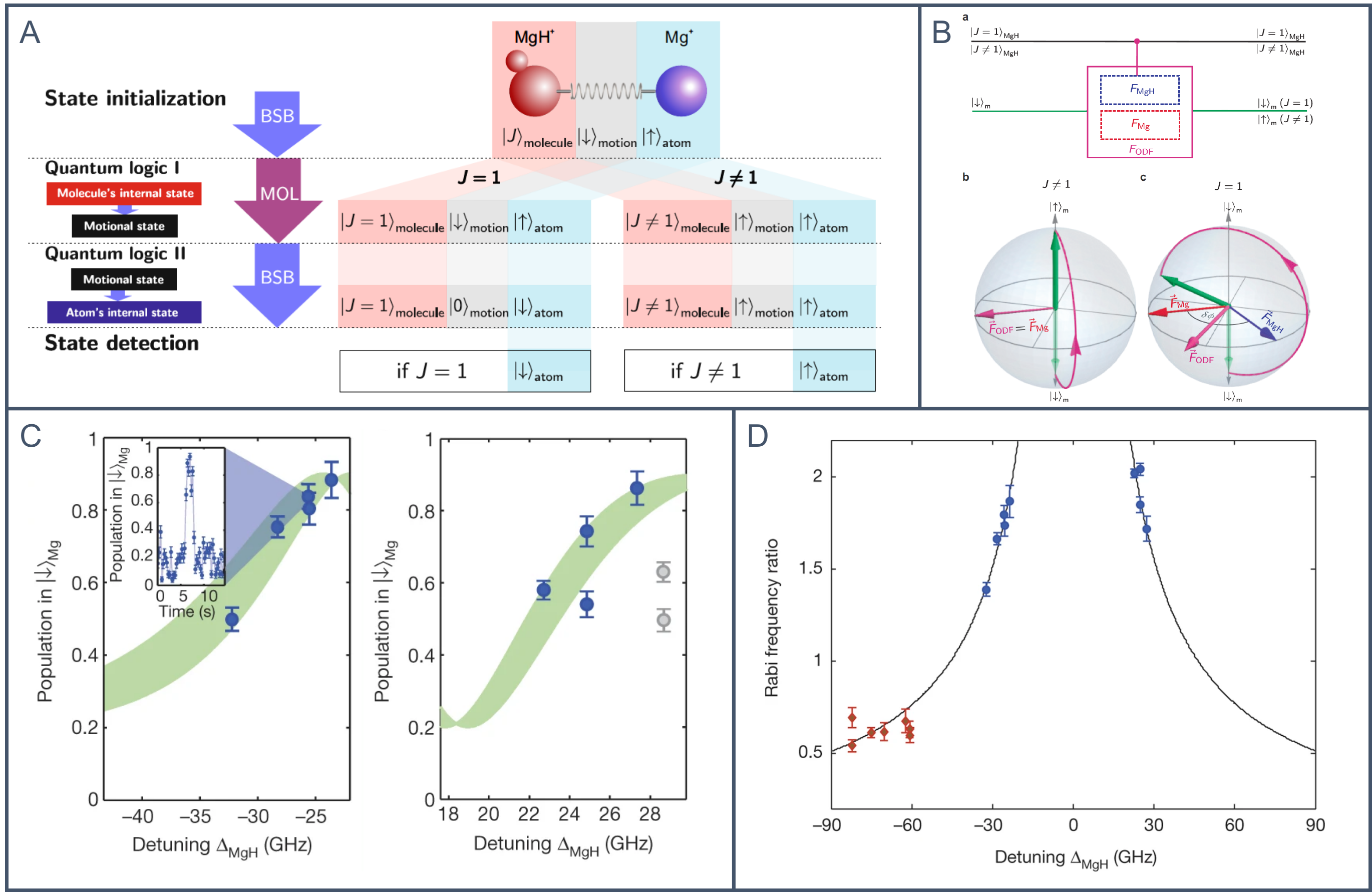}
    \caption[MgH$^+$ Spectroscopy]{{\bf A} Schematic representation of the quantum-logic spectroscopy implemented for the state detection of MgH$^+$ in Ref. \cite{wolf16a}. See text for details. Reproduced with permission from F. Wolf for \cite{wolf19a}. {\bf B}  Circuit diagram of the state-mapping process from the molecular to the motional state (panel a). A Bloch-sphere representation depicting the final motional qubit state for the $J \neq 1$ (panel b) and $J = 1$ (panel c) rotational states is also shown. 
    {\bf C} Theoretical estimation (green shaded region) and experimental observations (data points) for the population measurement in the ground state $\ket{\downarrow}_{\text{Mg}}$ of the Mg$^+$ ion provided that the molecule is in the $J = 1$ rotational state, after the implementation of the state-dependent force. 
    {\bf D} Mapping of a rovibronic transition in the MgH$^+$ ion by application of the non-destructive state detection technique. Panels B, C, and D have been adapted by permission from the Springer Nature Customer Service GmbH: Springer Nature for Wolf et al., \cite{wolf16a}, Copyright 2016.}
    \label{fig:MgH_Spec}
\end{figure*}

Fig. \ref{fig:MgH_Spec} shows a complementary approach to molecular-state identification in the presence of a spurious detuning-independent ODF experienced by the atomic ion. In these experiments \cite{wolf16a}, which represented the first quantum-logic spectroscopy performed on a molecular ion, a specific motional state was engineered on a Mg$^+$--\,MgH$^+$ two-ion chain. An ODF was then implemented in order to map the state of the molecular ion onto the motion which was subsequently detected by sideband measurements on the atomic ion. The experimental sequence is shown in Fig. \ref{fig:MgH_Spec}A. Here, both the IP and the OP motional modes of the two-ion chain were cooled to the ground state by resolved-sideband cooling \cite{wan15a}. In a first step, the motion was initialized in a state \ket{\downarrow}$_{\text{motion}} = \ket{1}_{\text{IP}}\ket{0}_{\text{OP}}$ corresponding to one quantum of excitation in the IP mode. The atomic ion was prepared in an excited state \ket{\uparrow}$_{\text{atom}}$. An ODF was then applied using a 1D optical lattice with parameters chosen such that the motional state of the two-ion crystal was changed to \ket{\uparrow}$_{\text{motion}} = \ket{0}_{\text{IP}}\ket{1}_{\text{OP}}$ by the ODF acting solely on the atom in case the molecular ion was not in the $J = 1$ rotational state (Fig. \ref{fig:MgH_Spec}Bb). By contrast, if the MgH$^+$ ion was in the targeted rotational state, the Mg$^+$--\,MgH$^+$ two-ion chain experienced an additional state-dependent ODF due to the molecular ion resulting in a different dynamics on the Bloch sphere of the motional qubit (Fig. \ref{fig:MgH_Spec}Bc). For suitably chosen experimental parameters, the protocol corresponds to a CNOT logic gate performed on the motional qubit with the molecular state as the control allowing to detect the $J = 1$ rotational state in the ground vibronic manifold.
In a subsequent step, a sideband measurement on the atomic ion was used to detect the motional state of the two-ion chain resulting in the determination of the molecular state. Fig. \ref{fig:MgH_Spec}C shows the probability of exciting the sideband on the Mg$^+$ ion after the ODF pulse as a function of the detuning, $\Delta_{\text{MgH}}$, of the lattice laser frequency on both the red and blue side of a molecular transition starting from the $J=1$ rotational state. The theoretical estimates, including the measurement uncertainties, are given by the green areas and the experimental realisations are represented by the data points. By determining the strength of the ODF acting on the molecular ion, a spectroscopic measurement on the relevant molecular transition was performed as shown in Fig. \ref{fig:MgH_Spec}D. 

\begin{figure*}[tp!] 
    \centering
    \includegraphics[width=1.0\linewidth,trim={0cm 0cm 0cm 0cm},clip]{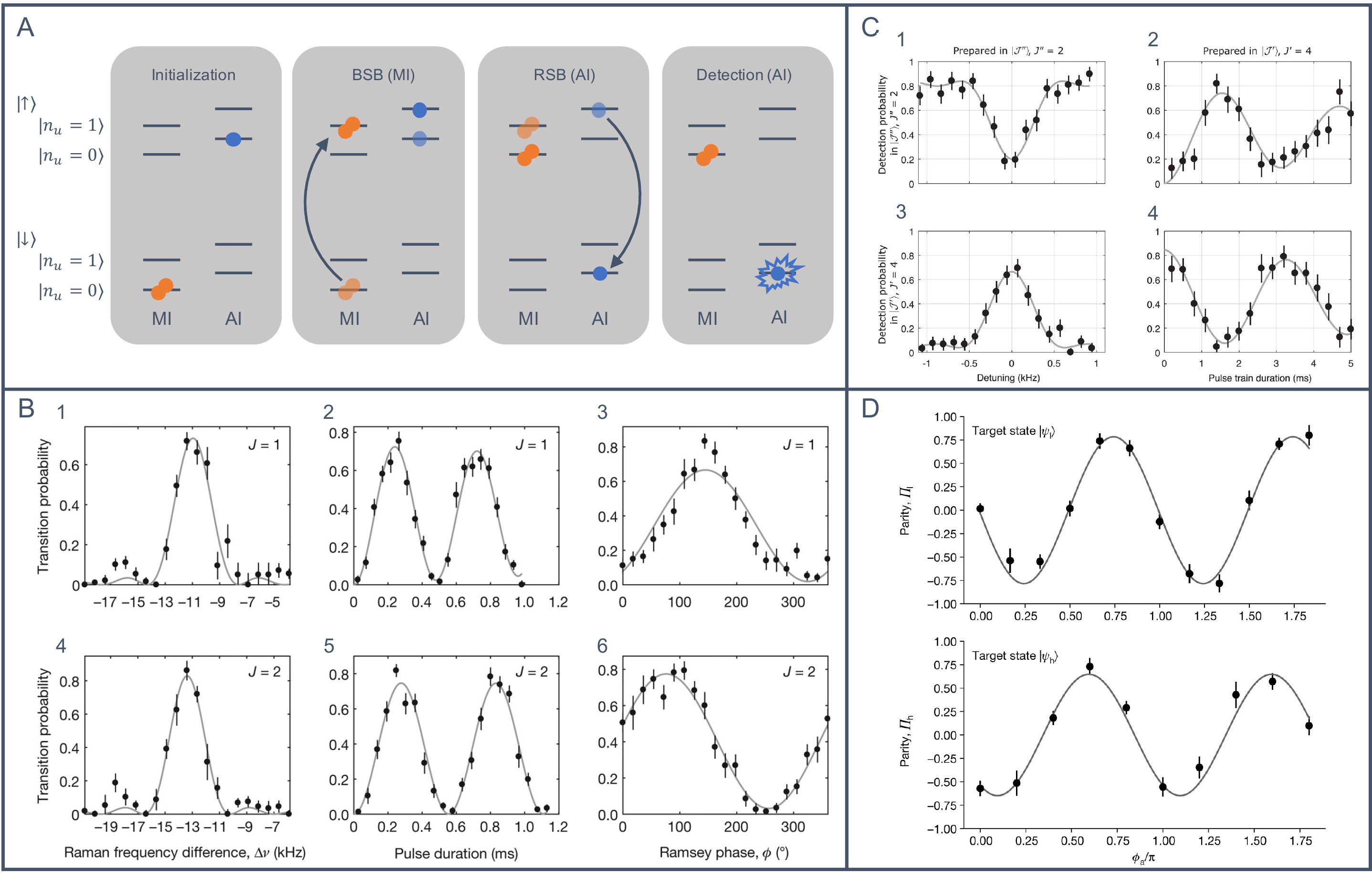}
    \caption[CaH$^+$ Spectroscopy]{Non-destructive state determination and coherent manipulation of CaH$^+$ ions. {\bf A} Schematic of the quantum-logic technique implemented for molecular state detection and projective preparation in Ref. \cite{chou17a} (MI=molecular ion, AI=atomic ion). See text for details
    \footnote{The definition of a red sideband as indicated in the third panel of Fig. \ref{fig:CaH_Spec}A adopted by the authors in Ref. \cite{chou17a} differs from the one used throughout this article (for example see Fig. \ref{fig:IonTrap_Basic}A where such a transition would be the labelled as a blue sideband).}.
    {\bf B} Coherent spectroscopy and manipulation on the $J = 1$ (panels 1, 2 and 3) and $J = 2$ (panels 4, 5 and 6) rotational states of the CaH$^+$ ion on transitions of the form $\ket{J, -J-1/2, -} \leftrightarrow \ket{J, -J+1/2, -}$. Frequency scans are shown in panels 1 and 4,  Rabi flops are depicted in panels 2 and 5, Ramsey fringes as a function of the relative phase between two $\pi/2$ pulses with a wait time of 15 ms are shown in panels 3 and 6. Reprinted by permission from the Springer Nature Customer Service GmbH: Springer Nature for Chou et al., \cite{chou17a}, Copyright 2017.
    {\bf C} High-resolution frequency spectra (panels 1 and 3) and Rabi flops (panels 2 and 4) driven by an optical frequency comb for $\Delta  J = 2$ rotational transitions in a single CaH$^+$ ion. Republished with permission of American Association for the Advancement of Science, from Chou et al., \cite{chou20a}, 367(6485), 2020; permission conveyed through Copyright Clearance Center, Inc.
    {\bf D} Parity measurements of entangled states between an atom and a low-frequency ($\ket{\psi_\text{l}}$) as well as a high-frequency ($\ket{\psi_\text{h}}$) molecular qubit. Reprinted by permission from the Springer Nature Customer Service GmbH: Springer Nature for Lin et al., \cite{lin20a}, Copyright 2020.}
    \label{fig:CaH_Spec}
\end{figure*}


\subsection{Molecular quantum logic spectroscopy using resolved-sideband Raman transitions}

A different approach to molecular-state preparation and detection was implemented by Chou et al. \cite{chou17a} in CaH$^+$ (see Fig. \ref{fig:CaH_Spec}). As in the previously discussed protocols, the experiment was started by cooling the shared motion of a \Ca{}--\,CaH$^+$ two-ion chain and preparing the \Ca{} ion in an excited state. In the room-temperature environment of the experiment in Ref. \cite{chou17a}, the molecules were prepared in their electronic and vibrational ground state and preparation of a specific rotational state relied on blackbody-assisted population redistribution. Raman laser beams were then applied in order to drive specific BSB transitions in the molecule such that a change in the molecular state was accompanied by an excitation of the shared motion of the two-ion chain (Fig. \ref{fig:CaH_Spec}A). The excited motion was subsequently detected by driving a red sideband on the atomic ion followed by fluorescence detection leading to projection of the molecule into a specific quantum state by the measurement. By allowing sufficient time for the molecule to re-equilibrate with the blackbody environment, Raman spectra of several molecular transitions could be recorded by sideband measurements on the atomic ion.

Since the times for population redistribution of the rotational states by BBR were long (about 100 ms to more than 2 s at room temperatures for $J<8$) compared to the time required to drive transitions ($<5$ ms), the population of the rotational states could be accumulated in specific magnetic sublevels by optical pumping and the targeted quantum states could then be coherently manipulated. The eigenstates of CaH$^+$ are labelled by the quantum numbers \ket{J, m, \xi}, where $J=0,1,2,3,..$ is the rotational angular momentum quantum number in this case, $m \in \{ -J-1/2, -J+1/2, J-1/2, J+1/2 \}$ denotes the sum of the components of the rotational angular momentum and the proton spin along the magnetic field direction. $\xi \in \{+,-\}$ indicates the relative sign of a superposition of product states with the same $m$ but opposite proton spin. Fig. \ref{fig:CaH_Spec}B depicts coherent spectroscopy and operations on the molecular ion on the $\ket{J, -J-1/2, -} \leftrightarrow \ket{J, -J+1/2, -}$ transitions for $J = 1$ (panels 1, 2 and 3) and $J = 2$ (panels 4, 5 and 6). 

Based on these state-detection and preparation techniques, a high-resolution rotational spectrum of a single CaH$^+$ molecule was recorded with a linewidth $<1$ kHz and an accuracy of $\sim1$ part per billion \cite{chou20a}. The spectra were obtained by coherently driving stimulated-Raman rotational transitions using an optical frequency comb (Fig. \ref{fig:CaH_Spec}C). Here, the molecule was prepared in the $\ket{\mathcal{J}''} = \ket{2, -5/2, -}$ state followed by probing the $\ket{\mathcal{J'}} \leftarrow \ket{\mathcal{J}''}$ transition where $\ket{\mathcal{J'}} = \ket{4, -7/2, -}$. The probabilities of the molecule being in either states as a function of the Raman difference frequency from the resonance is shown in panels 1 and 3. When prepared in the $\ket{\mathcal{J'}}$ state, the coherent Rabi flops to and from the $\ket{\mathcal{J''}}$ state are shown in panels 2 and 4.

Moreover, by applying tailored pulse sequences on the the \Ca{} and and CaH$^+$ ion, a quantum entanglement was generated between the atom and the molecule \cite{lin20a}. Fig. \ref{fig:CaH_Spec}D shows parity measurements of entangled states of the atom and a low-frequency ($\ket{\psi_\text{l}}$) as well as a high-frequency ($\ket{\psi_\text{h}}$) molecular qubit. The state parities, $\mathit{\Pi}$, were measured as a function of the phase of an `analysis' pulse, $\phi_\text{a}$, and oscillate as $C \cos(2\phi_\text{a} + \phi_0)$ where $\phi_0$ is an offset in phase and $C\geq0$ is the observed contrast. The fidelities for the generation of the low-frequency and high-frequency entangled states in this experiment were estimated to be $F_\text{l} = 0.87(3)$ and $F_\text{h} = 0.76(3)$ indicating bipartite entanglement for both cases \cite{leibfried05a}. 

\section{Outlook on future developments and conclusions}

In addition to the already realised experiments discussed above, alternative quantum-logic protocols have been proposed for molecular ions. Mur-Petit et al. \cite{mur-petit12a} have suggested temperature-independent geometric-phase gates based on state-dependent forces for non-destructive molecular-state detection. These operations are insensitive to the motional state of a trapped atom-molecule two-ion crystal and can thus be applied to Doppler cooled ions possibly leading to simpler and faster experimental implementations. 

Loh et al. \cite{loh14a} have proposed a state-detection technique relying on state-dependent magnetic $g$ factors which lead to distinct Zeeman splittings for electronic, rotational and hyperfine quantum states of a molecule. Such a Zeeman-splitting-assisted quantum logic spectroscopy (ZSQLS) scheme would employ lasers far detuned from one-photon transitions in the molecule to drive Raman transitions between Zeeman sublevels without scattering photons and thus preserve the quantum state of the molecular ion. Apolar molecules like O$_2^+$ and polar molecules like SO$^+$, CO$^+$ and SiO$^+$ could be probed by the ZSQLS technique under suitable experimental conditions \cite{loh14a}. 

As an extension of the CaH$^+$ state-detection scheme implemented by Chou et. al \cite{chou17a}, Wolf et al. \cite{wolf20a} proposed a quantum-logic technique for non-destructive state detection of O$_2^+$ ions that would allow for amplification of the state detection signal which is particularly important if single-shot readout of the atomic ion is not possible for technical reasons. The scheme employs bichromatic far-detuned Raman lasers to drive red and blue sideband transitions between selected Zeeman states simultaneously. For specific rotational states, the phase of the bichromatic drive can be chosen such that a Schr{\"o}dinger cat state of the form $\ket{\psi}_{\text{SC}} = \ket{+}\ket{\alpha} \pm \ket{-}\ket{-\alpha}$ is produced, where $\ket{\pm} = (\ket{1/2}\pm\ket{3/2})/\sqrt{2}$ and $\ket{\pm\alpha}$ denotes a coherent state with complex amplitude $\pm \alpha$. By detecting the depletion of the motional ground state due to the emergence of the Schr{\"o}dinger cat state after application of the Raman pulses, the successful drive of the transition can be detected on the atomic ion making it possible to infer the molecular state. 

Another series of proposals explored the potential of using the dipole moments of trapped polar molecular ions as a resource for quantum logic, either by capitalising on direct dipole-dipole couplings between the molecules \cite{hudson18a} or the coupling of the dipoles to the normal modes of a Coulomb crystal \cite{campbell20a, mills20a}. The latter approach offers possibilities for implementing protocols which do not require a direct optical addressing of the molecules and their cooling to the motional ground state. Proposals for further developments of the dipole-phonon toolbox include the application of microwave fields to address both the internal states and the motions of the trapped polar ions \cite{hudson21a}.

A variant of quantum-logic spectroscopy makes use of the recoil imparted by the absorption or emission of photons to the motion of ground-state cooled ions. The resulting change of motional state can then be detected by the usual sideband-spectroscopy techniques. This approach has proven useful in the precise measurement of strong, dipole-allowed transitions in atomic ions \cite{wan14a}. Extensions of this method to infrared spectroscopy \cite{clausen20a} and to probing the vibrational dynamics dynamics of trapped molecular ions \cite{schindler19a} have recently been proposed. 

The new coherent methods for molecular ions discussed are not only of interest for applications like quantum science and precision spectroscopy, but also have a considerable potential in completely different realms such as collision studies and chemistry. Proof-of-principle experiments were presented in Refs. \cite{sinhal20a, najafian20a} in which changes of the internal state \cite{sinhal20a} and of the chemical identity \cite{najafian20a} of single N$_2^+$ ions were tracked as a consequence of collisions demonstrating the utility of the present methods as non-invasive probes for investigating inelastic and reactive collisions of molecular ions. 

In parallel to the methodological advances outlined above, new technological developments are addressing an ever growing variety of molecular systems. Recently laser cooling and optical-pumping schemes for molecular cations like BH$^+$ \cite{nguyen11a}, SiO$^+$ \cite{stollenwerk17a, li19a}, TeH$^+$ \cite{stollenwerk18a}, AlCl$^+$ and AlF$^+$ \cite{kang17a}, and anions like C$_2^-$ \cite{yzombard15a} have been explored theoretically. Apart from the systems already used in current experiments, i.e., MgH$^+$, CaH$^+$ and N$_2^+$, quantum-logic techniques for molecular ions like H$_2^+$ and its isotopologues \cite{schiller05a, karr14a,schiller14a}, O$_2^+$ \cite{hanneke16a, wolf20a} and polyatomics such as cyanoacetylene \cite{schindler19a} and propanediol \cite{patterson18a} haven been explored in various proposals. Clearly, with the vast number of molecular systems available, the possibilities are virtually endless.

The coherent experimental tools available for trapped molecular ions have now reached a similar level of sophistication as the ones employed for atomic ions. Although the field of molecular-ion quantum technologies is still in its infancy, the presently realised experiments as well as the current proposals for future advances discussed in this review illustrate the immense potential for the quantum control and manipulation of molecular ions and the ensuing applications in precision spectroscopy, quantum science and even chemistry. 

\section*{Acknowledgements}

Our work is funded by the Swiss National Science Foundation (SNSF) under project grants CRSII5\_183579 and 200021\_204123. It is also supported as a part of NCCR QSIT, a National Centre of Competence in Research, funded by the Swiss National Science Foundation (grant number 51NF40-185902).


%

\end{document}